\documentclass[runningheads]{llncs}

\newcommand{\minarets}{\textsf{Minarets}\xspace}
\newcommand{\barrow}{\textcolor{blue}{->}}
\newcommand{\evt}{\textcolor{blue}{eventually!}\xspace}
\newcommand{\eg}{\textcolor{blue}{EG}\xspace}
\newcommand{\ag}{\textcolor{blue}{AG}\xspace}
\newcommand{\ef}{\textcolor{blue}{EF}\xspace}

\newcommand{\assert}{\textcolor{blue}{assert}\xspace}
\newcommand{\propertyb}{\textcolor{blue}{property}\xspace}
\newcommand{\for}{\textcolor{blue}{for}\xspace}
\newcommand{\inn}{\textcolor{blue}{in}\xspace}
\newcommand{\e}{\textcolor{blue}{and}\xspace}
\newcommand{\orr}{\textcolor{blue}{||}\xspace}
\newcommand{\bool}{\textcolor{blue}{boolean}\xspace}
\newcommand{\clock}{\textcolor{blue}{clock}\xspace}
\newcommand{\integer}{\textcolor{blue}{integer}\xspace}
\newcommand{\always}{\textcolor{blue}{always}\xspace}

\usepackage[english]{babel}   
\usepackage{ucs}
\usepackage[utf8x]{inputenc}
\usepackage[T1]{fontenc}
\usepackage{pslatex}
\usepackage{graphicx}
\usepackage{xspace}
\usepackage{float}
\usepackage{color}
\usepackage{alltt}
\usepackage{cite}
\usepackage{soul} 
\usepackage{mathrsfs} 
\usepackage{listings}
\usepackage{mdframed}
\usepackage{xcolor}
\usepackage{comment}
\usepackage{booktabs}
\usepackage{centernot}
\usepackage{threeparttable}
\usepackage{soul}
\usepackage{caption}
\usepackage{url}
\usepackage{amssymb}
\usepackage{graphics}
\usepackage{amsmath,bm,times}
\usepackage{mathrsfs}
\usepackage{mathtools}
\usepackage{caption}
\usepackage{capt-of}
\usepackage{dblfloatfix} 
\usepackage[many]{tcolorbox}

\usepackage[colorlinks=blue, linktocpage=false]{hyperref}
\title{Modeling and Analysis of the Landing Gear System with the Generalized Contracts}
\author{A. Abdelkader Khouass\inst{1,2} \and  J. Christian Attiogbé
	\inst{1} \and Mohamed Messabihi\inst{2} \and Abdelkrim Benamar \inst{2}}
\institute{University of Nantes, LS2N CNRS UMR 6004, France \\ 
	\email{christian.attiogbe@univ-nantes.fr}\\\email{abderrahmaneabdelkader.khouass@univ-tlemcen.dz } \and
	University of Tlemcen, LRIT, Algeria\\	
	 \email{mohamedelhabib.messabihi@univ-tlemcen.dz }\\ \email{Abdelkrim.Benamar@univ-tlemcen.dz } }
\newtcolorbox{minimal}{%
	sharp corners, colback=white, colframe=black, notitle,
	before skip=1cm, after skip=1cm}
\begin{document}
\maketitle
\begin{abstract}
Nowadays, there are several complex systems in different sectors such as aviation, air traffic control ...etc.
These systems do not have a precise perimeter, they are open and made of various specific components built with different languages and environments.
The modeling, assembly and analysis of such open and complex heterogeneous systems are challenges in software engineering.
This paper describes how the \textit{\minarets} method decreases the difficulty of modeling, composition and analysis of the well known case study of the landing gear system.  
The method consists in: equipping individual components with \textit{generalized contracts} that integrate various \textit{facets} related to different concerns, composing these components according to their facets and verifying the resulting system with respect to the involved facets as well. 
The proposed method may be used or extended to cover more facets, and by strengthening assistance tool through proactive aspects in modeling, composing multi-facets contracts and finally the verification of the heterogeneous systems.
\end{abstract}

\vspace{1cm}
\section{Introduction}\label{section:intro}

\noindent Several system companies in different sectors (air traffic control, nuclear power plants, railway, autonomous vehicles, etc.) are facing an exponentially increasing complexity of their systems; their systems are assembled by a huge number of hardware/software components. Their complexity forces one to have a wide variety of heterogeneous components, due to the requirements. 

 Furthermore, there exist various types of composition to compose the concerned components, for instance, the parallel composition, sequential composition and additive composition; in addition, the components may cover different \textit{facets} Therefore, if the integration of components into a global system is not mastered, it may generate considerable time losses and overcharges due to inconsistency of requirements, incompatibility of meaning and properties, late detection of composition errors, etc.
For these reasons, the modeling, the composition and formal analysis of such  heterogeneous systems are challenging. The use of efficient methods and techniques is required to face these challenges.

We aim at studying and alleviating the difficulties of practical modeling and integration of heterogeneous components.  

This paper presents a simplified version of the landing gear system described in~\cite{landing_boniol_details}. We describe the main components to show how we decrease the difficulty of the modeling and the analysis of the system with our \minarets method presented previously in "\textit{10th International Conference on Model and Data Engineering}"~\cite{khouass2021}; we also detail the composition of \textit{normalized components} and finally we compare our approach of modeling and verifying of the landing gear system with the related approaches. The challenge is not to do the complete case study as required in the specification.

In the previous work \cite{khouass2021} we proposed a method (named "ModelINg And veRifying heterogeneous sysTems with contractS" (\textsf{Minarets})), which consists in modeling and verifying a system with the concept of \textit{generalized contracts}. The contract is generalized in the sense that it will allow one to manage the interaction with the components through \textit{given facets}: the properties of the environment, the properties of the concerned components, the communication constraints, non-functional properties (quality of service for example), etc. 

We focus on the generalized contracts not only in the composition of the individual components, but in the verification as well; this reduce the complexity of the analysis of heterogeneous systems; moreover, the structuring of contracts with facets and priority of properties, makes it possible to decrease the difficulty of the heterogeneous systems assembly and to be more efficient during the verification.


The rest of the article is structured as follows.
Section \ref{section: challenges} is dedicated to the challenges of heterogeneous systems,
Section \ref{section:Methodology} introduces the materials and the modeling and verification methodology.
In Section \ref{section: case study} we present the landing gear system and we model and analyse it with our \minarets method.
Section \ref{section:related work} provides an overview of related work, and finally, Section \ref{section:conclusion} gives conclusions and future work.\\

\section{Challenges of heterogeneous systems }\label{section: challenges}
\noindent Among other challenges of heterogeneous systems there are composability, composition and compositionality.\\
\noindent\paragraph{Composition}
Because a component requires a functionality offered by another one, it can not be directly usable, i.e. the call of one of its functionalities will not work if the components are not bound in advance. The composition of components consists in binding them in~an interaction relationship \cite{messabihi_these}.\\  

\noindent\paragraph{Composability}
Composability means stability of component structures across integration. That means given components can be composed as they are. A component’s structure is not affected during the system construction process, i.e., even after composing the different components together, their individual structures are preserved. \\

\noindent\paragraph{Compositionality}
Compositionality denotes the inferring of global system properties from the local properties of the components. An example is inferring global deadlock-freedom from the deadlock freedom of the individual components\cite{sif2014}.\\
\section{Modeling and verification using generalized contracts}
\label{section:Methodology}
An issue to be solved for heterogeneous systems is that, the interfaces of involved components should be composable. For the sake of simplicity of the composition, we adopt the well-researched concept of \textit{contract} which is therefore extended for the purpose of mastering heterogeneity of components interfaces. Moreover, for a given system we will assume \textit{agreed-upon facets} such as data, functionality, time, safety, etc.

\subsection{Materials}
In the proposed method, we extend the traditional Assume-Guarantee (A-G) contract with the purpose of mastering the modeling and verification of complex and heterogeneous systems.

\begin{definition}[Contract] \label{contract}
	A contract C of a given component is a pair (A, G) of predicates, where A is the assumption, and G, the guarantee. The assumption must be respected at the entry of the component and the guarantee must be fulfilled at the exit of the component \cite{Mey1992}.   
\end{definition}
We extend the contract to face our challenges.
\begin{definition}[Generalized contract]\label{GC}
	A \textit{generalized contract} is a multi-faceted Assume-Guarantee contract. It is an extension of contract, structured on the one hand with its assume and guarantee parts, and structured on the other hand according to different clearly identified and agreed-upon facets (data, functionality, time, security, quality, etc.) in its assume or guarantee.
\end{definition}

The generalized contract will be layered to facilitate properties analysis.
Every facet will have a priority which is a natural number $(n \in \mathbb{N})$. Therefore, an analysis of a facet may be done prior to another facet. A layer is denoted by a facet $F$ and its priority $n$; thus, we will have a couple $(Facet, n)$ for each layer; we will write simply $Facet[n]$.  \\

\noindent An example of a generalized contract is as~follows:\\
(A, G) = (\{(\textbf{DATA}[1], Predicate),\\
\hspace*{1.55cm}(\textbf{TIME}[2], Predicate), ...\},\\
\hspace*{1.4cm}\{(\textbf{TIME}[2], Predicate),\\
\hspace*{1.55cm}(\textbf{SECURITY}[3], Predicate), ...\})\\
\begin{definition}[Normalized component]\label{wsc}
	A well-structured or normalised component is a component equipped with a generalized contract, acting as its interface with other components. 
\end{definition}

Normalising a component $C_i$ consists in transforming $C_i$ into a component equipped with a generalized contract.
A multi-faceted A-G contract will be expressed in a wide purpose expressive language.

In this work we chose  the PSL language  \cite{PSL} to specify contracts. PSL is a formal  language for specifying properties and behaviour of systems. It is an extension of the Linear Temporal Logic (LTL) and the Computation Tree Logic (CTL). PSL could be used as input for formal verification, formal analysis, simulation and hybrid verification tools. PSL improves communication between designers, architects and verification engineers. We use the ALDEC Active-HDL \footnote{https://www.aldec.com/en/products/fpga$\_$simulation/active-hdl} tool that supports PSL.

For experimentation purpose, we use ProMeLa and SPIN~\cite{spinprom}, the model checker UPPAAL\cite{uppaal} to model and verify components. SPIN is an automated model checker which supports parallel system verification of processes described with its input  PROtocol MEta LAnguage (ProMeLa).
\subsection{Outline of the proposed method: \minarets}
\begin{figure}[!h]
	\centerline{\includegraphics[scale=0.32]{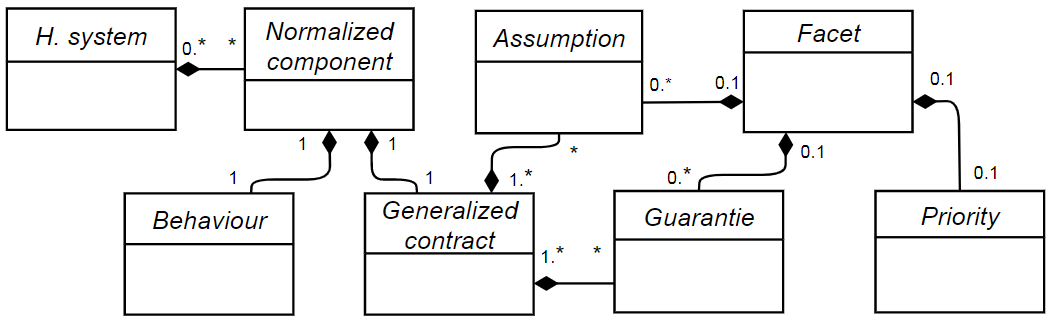}}
	\captionsetup{justification=centering} \caption{Meta-model of a system with normalized~components}
	\label{figure:meta_model}
\end{figure}

The working hypothesis is that a heterogeneous system should be an assembly of \textit{normalized components} (see Def. \ref{wsc}). This is summarized in Figure \ref{figure:meta_model} with the block definition diagrams of SysML\footnote{System Modeling Language \cite{sysml}}. Each component has a generalized contract and a behaviour expressed with \textit{labelled transition system} (LTS); after that, for the sake of the verification efficiency, we may add a priority to each facet.   \\

\noindent The method that we propose (\textsf{Minarets}) consists in,  given a set of appropriately selected or predefined elementary components, normalizing these input components prior to their composition, building a global heterogeneous system, and finally analysing this global system with respect to the required properties.

For this purpose there are many issues to be solved:\\
\textit{i)} Elementary components are from various languages and cover different facets, a pragmatic means of composition is required. We consider PSL as a wide purpose expressive language to describe  generalized contracts. Each component will be manipulated through its \textit{generalized contract} (see Def. \ref{GC}) written in an appropriate language.\\ 
\textit{ii)} Global properties are heterogeneous; they should be clearly expressed, integrated and analysed; they will be expressed with a wide purpose language such as PSL; we will decompose them according to the identified \newline agreed-upon facets and  spread them along the analysis of composed components.\\
\textit{iii)} Composition of elementary components should preserve their local requirements and should also be weakened or strengthened with respect to global-level properties. For instance, some facets required by an elementary component could be unnecessary for a given global assembly, or some facets required at a global assembly may be strengthened at a component level. \\
\textit{iv)} Global properties require heterogeneous formal analysis tools; this generates complexity. We choose to separate the concerns, so as to target various tools and try to ensure the global consistency.\\
\textit{v)} Behaviours of components should be composable.\\

\begin{figure*}[!h]
	\begin{center}
		{\includegraphics[scale=0.56]{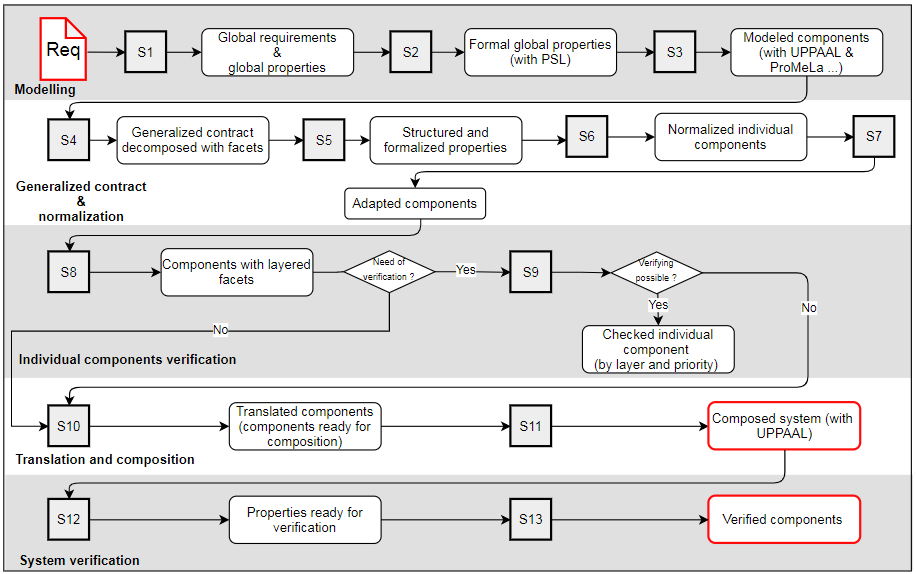}}
		\caption{The \textsf{Minarets} process chart}
		\label{figure:summarized_steps}
	\end{center}
\end{figure*}

The working flow of the \textsf{Minarets} method is depicted in Figure \ref{figure:summarized_steps} with 13 steps (S1, S2,~$\cdots$~S13). For the sake of brevity, we  present the different steps of our method along a case study; more details can be found in our research reports \cite{minarets}.  

The \textsf{Minarets} method integrates solutions to these issues, as we will present in the sequel. 
We adopt a correct-by-construction approach for the assembly of components. Therefore, local compositions should preserve required properties of components.
In the same way, global properties may impact the components; therefore, global properties are decomposed and propagated through the used components when necessary.
\subsection{Notations}
We denote by $C_{i}$=$(GC_{i},B_{i})$ 
a normalized component; where $GC_{i}$ is the generalized contract of  $C_{i}$ and $B_{i}$ is the behaviour of $ C_{i} $. The behaviour $B_i$ is represented with a LTS. Let consider $\mathscr{C}$ as a set of components, $\mathcal{P}$ a set of properties, $\mathcal{F}$ a set of facets.\\ 

\noindent A generalized contract GC is denoted by $GC$=$(A,G)$ where $A$ is the assumption and $G$ is the guarantee. The assumption $A$ or the guarantee $G$ is a set of properties $P_k \in \mathcal{P}$; where each property has a facet $F_{i}$; $F_i \in \mathcal{F}$ and each facet $F_i$ has a priority $n_j$; then,\\
$A$=$\{(F_{i}[n_j], P_{k})\}$, $G$=$\{(F_{i}[n_j],P_{k})\}$.  
\section{Case Study}\label{section: case study}
In this section we introduce the landing gear system, after that, we show how we contribute to decrease the difficulty of its modeling and analysis using our \minarets method.\\
\subsection{General Description of the Landing Gear System } 
\begin{figure}[H]
	\centering
	\includegraphics[scale=0.41]{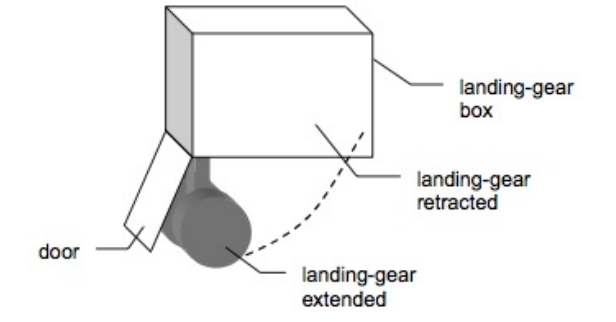}
	\caption{Simplified view of the landing system\cite{landing_boniol_details} }
	\label{figure:simplified_landing_system}
\end{figure}

\begin{figure}[H]
	\centering
	\includegraphics[scale=0.60]{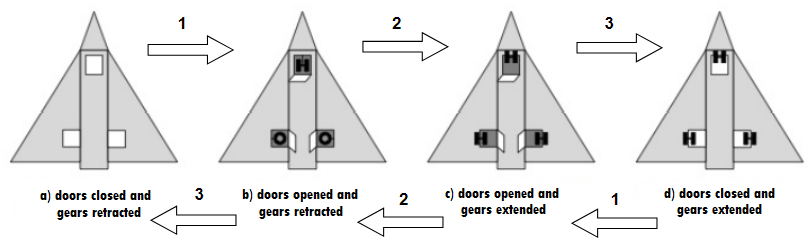}
	\caption{The landing/taking off sequence \cite{landing_boniol_details} }
	\label{figure:landing_sequence}
\end{figure}

The landing system \cite{landing_boniol_details} is in charge of operating the landing gear and the associated doors.
The landing system is made up of 3 landing sets:  front, right and left.
Each landing set contains a door, a landing gear and associated hydraulic cylinders. In this case study we consider only the front door and the front landing gear without taking into consideration the hydraulic cylinders; furthermore, we consider only one sensor to prevent the state of each mechanical parts instead of 3 sensors (triplicated) as mentioned in the original version \cite{landing_boniol_details}. They use 3 sensors where each sensor send a value, the three values must be the same, otherwise, if one value of the three ones is different it will be eliminated, they consider only the two remaining values, after that, if these two values are not equal, it is a failure. A simplified schema of a landing system is shown in Figure~\ref{figure:simplified_landing_system}.
\noindent The basic landing sequence is depicted in the Figure \ref{figure:landing_sequence}; first, open the landing gearbox doors; second, extend the gears; third, shut the doors.
 After the take-off, the basic retraction sequence is: open the doors, retract the landing gear and shut the doors. 
\subsubsection{Architecture of the Landing Gear System}\leavevmode \\
As depicted in the Figure \ref{figure:architecture}, the landing gear system is made up of 2 principal parts:\\
\begin{minipage}{0.55\textwidth}\raggedright
The mechanical part: door and the landing gear.\\
The cockpit: the up/down actuator and the lights.\\

\subsubsection{The cockpit}
To control the retraction and extension of the gears, an Up/ Down actuator is available to the pilot. When the actuator is switched to “Down”, the landing sequence is executed, when the actuator is switched to “Up” the retracting sequence is executed.
\end{minipage}
\begin{minipage}{0.45\textwidth}
	\begin{figure}[H]
		\centering
		\includegraphics[scale=0.15]{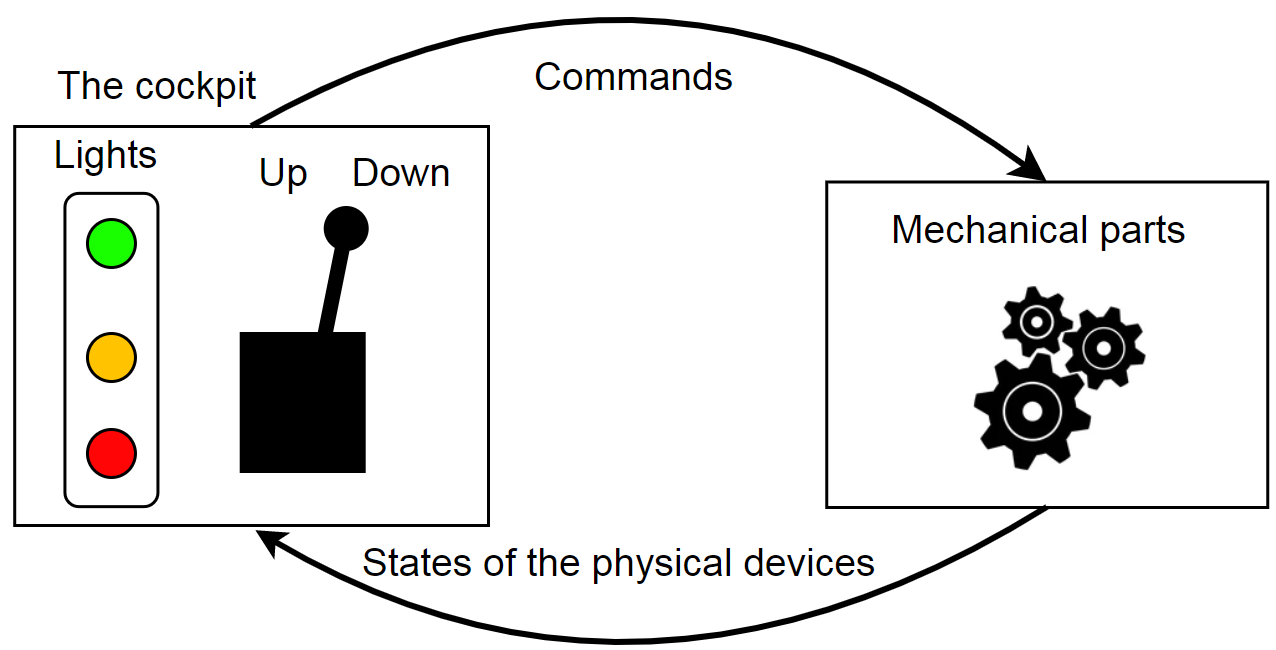}
		\caption{The Architecture of the system}
		\label{figure:architecture}
	\end{figure}
\end{minipage} \\ \leavevmode \\ \\
The pilot has a set of lights giving the current position of the gear, the door and the current status of the system to know if every thing work correctly or there is a  failure.\\
\begin{itemize}
	\item The locked gear in down position and the door closed in high position is represented by the green light.\\
	\item The gears or the door are manoeuvring, the orange light is on.\\
	\item The failure cases in the landing gear system are represented by the red light.  \\
	\item No lights are on when the gears are locked up and the door locked in high position.
\end{itemize}

\subsubsection{The Mechanical Parts}
\begin{figure}[t]
	\centering
	\includegraphics[scale=0.38]{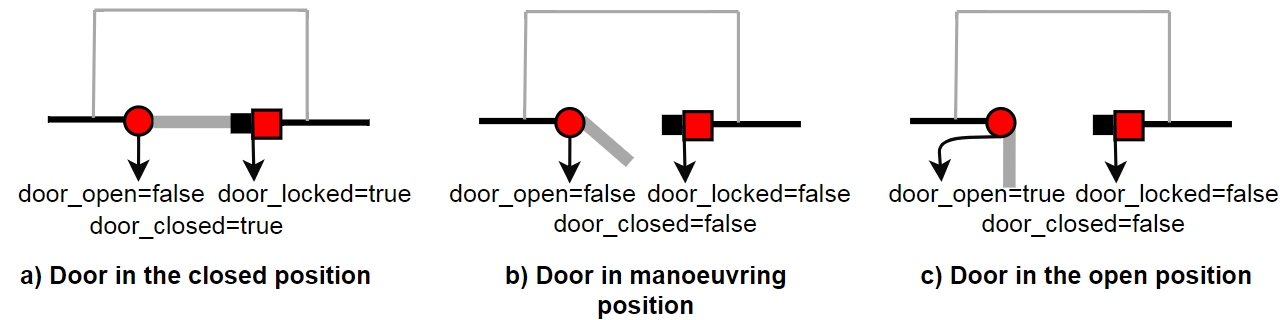}
	\caption{The different states of the door with its high latching}
	\label{figure:open_door}
\end{figure}

\begin{figure}[!h]
	\centering
	\includegraphics[scale=0.38]{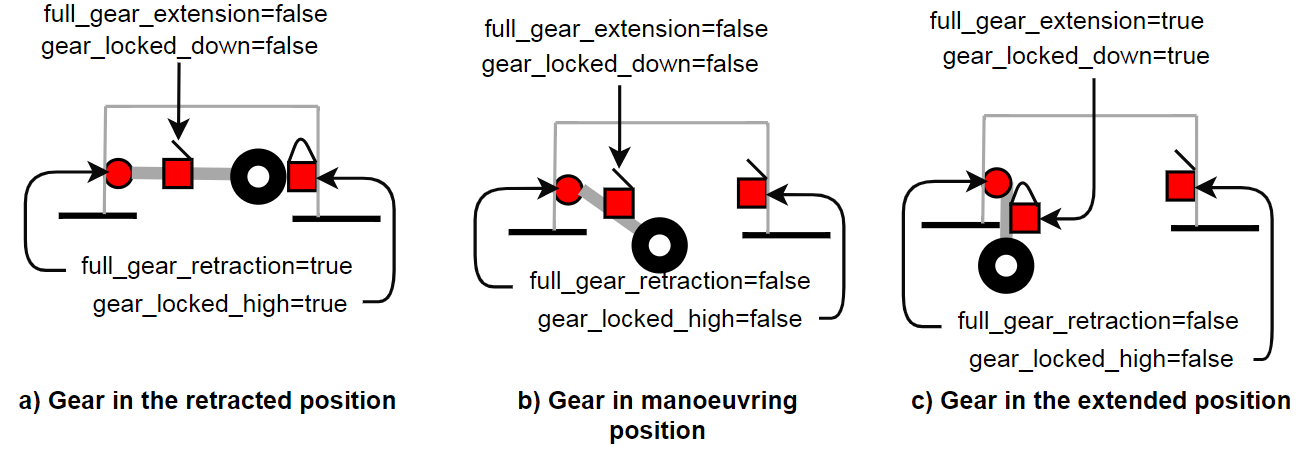}
	\caption{The Gear states with its high/down latching}
	\label{figure:open_gear}
\end{figure}

The system is composed from the landing gear, the doors, two latching of the gears down and up, the door latching and a set of sensors, as depicted in the Figures \ref{figure:open_door} \ref{figure:open_gear} .


\vspace{-\parskip}\vspace{-\parskip}

\subsection{Modeling and Verifying the Landing Gear System with MINARETS}
In the following, we present the LGS modeling and verification using our \minarets method.\\
\\ 
\textbf{Step 1 \textit{(modeling)}.} We express the informal global requirements and global properties with any desired language; then, we choose the natural language. The only goal here is to clearly state the requirements of the landing gear system.\\ \\
\noindent First, the initial state of the landing gear system is:\\
\noindent The door is closed and locked in high position.\\
The aircraft is in cruise mode, i.e. it is not in landing or take off position.\\
The gears are fully retracted and locked in high position.\\
The actuator position is up.\\
There is no light on in the pilot interface.  \\

\noindent \textit{The extension sequence}: if the system is in the initial state and the pilot operates the actuator to down, the system follows these actions:

\begin{enumerate}
	\item Unlock the door latching.
	\item Opening of the door (up to down position).
	\item Once the door is in open position, the gear will be in unlocked position.
	\item Extend the gear from high to down position.
	\item Lock the gear in down position. 
	\item Close the door (down to up position).
	\item Lock the door in high position. 
\end{enumerate}

We consider that the pilot could operate the actuator from down to up only after the end of the extension sequence.\\ 

\noindent \textit{Retraction sequence}: after the finishing of the extension sequence, i.e. the door is closed and locked in high position and the gears are locked in down position; the pilot could retract the gears by operating the actuator from down to up. These are the elementary actions of the system:
 \begin{enumerate}
 	\item Unlock the door latching.
 	\item Opening of the door (up to down position).
 	\item Once the door is in open position, the gear will be in unlocked position.
 	\item Retract the gear from down to high position.
 	\item Lock the gear in high position.
 	\item Close the door (down to up position).
 	\item Lock the door in high position. 
 \end{enumerate}  
\begin{table}[!hth]
		\centering\caption{The different time constraints of the gear and the door \cite{landing_boniol_details}}\label{tab:time_constraints}
		\begin{tabular}[t]{p{4cm} p{2cm} p{2cm}}
		\toprule
		\textbf{Duration (in seconds) of:}  &  \textbf{Front gear} & \textbf{Front door}\\
		\midrule
		Unlock in down position & 0.8 & - \\
		\midrule
		From down to high position & 1.6 & 1.2\\
		\midrule
		Lock in high position & 0.4  &  0.3\\
		\midrule	
		Unlock in high position & 0.8 & 0.4 \\
		\midrule
		From high to down position & 1.2 & 1.2\\
		\midrule
		Lock in down position & 0.4 & - \\
		\bottomrule
	\end{tabular}	
\end{table}			

\subsubsection{Time constraints} 
According to the table \ref{tab:time_constraints}, for instance, if the system is in the initial state and the pilot operates the actuator to down (he want to extend the gears), the overall operation of the full extension takes 5,5 seconds. First, the door unlocked in high position 0.4s; second, the door moving from high to down during 1,2s; after that, the gear unlocked in high position 0,8s; next, the gear move from high to down 1,2s; then, it will be locked in down position 0,4s; the door moving from down to high 1,2s and finally the door locked in high position 0,3s.\\ \\ 
\noindent \textbf{Global properties:} We specify the global properties as follows:\\
Note that we present "Pi bis" as a weaker version of Pi, the weaker version doesn't take into consideration the quantitative time, and we multiply the time value by 10 because UPPAAL doesn't support clocks with real values.\\\\

\noindent\textbf{/*----- (1) Combined properties $(P_i)$ -----*/}\\
\noindent
P1: In landing mode, eventually the gear must be fully extended and locked in down position.\\\\
P2: In cruise mode the gear must be retracted and the door is locked in high position (the cruise mode means that there is no ongoing landing or retraction operation) .\\\\
P2.1: In cruise mode the gear must be retracted. \\\\
P2.2: In cruise mode the door is locked in high position \\\\
P3: If the gear is manoeuvring from high to down, then the door must be completely~open.\\\\
P4: If the gear is manoeuvring from down to high, then the door must be completely~open.\\\\
P5: If the landing gear actuator has been pushed DOWN,  then eventually the door latching will be unlocked in 4s.\\\\
P5 bis: If the landing gear actuator has been pushed DOWN, then eventually the door latching will be seen unlocked.\\\\
P6: If the landing gear actuator is pushed down and stays DOWN, then eventually the door will be completely opened in 16 s.\\\\
P6 bis: If the landing gear actuator is pushed down and stays DOWN, then eventually the door will be completely open.\\\\
P7: The gear never manoeuvring from down to high and the door is not open.\\\\
P8: The gear never manoeuvring from high to down and the door is not open.\\\\
P9: If the landing gear actuator is DOWN and the door is completely open, then eventually the gear will be unlocked in high position in 8s.\\ \\
P9 bis: If the landing gear actuator stays DOWN and the door is completely open, then eventually the gear will be unlocked.\\ \\
P10: If the landing gear actuator is DOWN and the door is open and the gear is unlocked in high position, then eventually the gear will be extended in 20s after the door opening.\\ \\
P10 bis: If the landing gear actuator stays DOWN and the door is open and the gear is unlocked in high position, then eventually the gear will be seen extended.\\ \\
P11: If the landing gear actuator is DOWN and the gear is extended, then eventually the gear will be locked in down position in 24s after the door opening.\\   \\
P11 bis: If the landing gear actuator is DOWN and the gear is extended, then eventually the gear will be locked in down position.\\ \\
P12: If the actuator is down and the gear is locked in down position, then eventually the door will be closed and locked in high position in 55s after the first pushed down of the actuator (the overall extension operation takes 55s).\\\\
P12 bis: If the actuator is down and the gear is locked in down position, then eventually the door will be seen closed and locked in high position.\\\\
P13: After the take-off, if the gear is fully extended and the door is closed and the pilot push the actuator to up, the door will be open in 16s.\\ \\
P13 bis: After the take-off, if the gear is fully extended and the door is closed and the pilot push the actuator to up, the door will be seen open.\\ \\
P14: Always, After the take-off, if the the actuator stays up and the door is open and the gear is extended, then the gear will be retracted and locked in high position in 20s after the door opening.\\  \\
P14 bis: Always, After the take-off, if the the actuator stays up and the door is open and the gear is extended, then the gear will be seen retracted and locked in high position.\\    \\
P15: After the take-off, if the actuator stays up, and the gear is locked in high position, then the door will be closed in 59s after the first pushed up of the actuator (the overall retraction operation takes 59s).\\ \\
P15 bis: After the take-off, if the actuator stays up, and the gear is locked in high position, then the door will be seen closed.\\ \\
P16: If the actuator is down, and stays down, the retraction sequence is never observed.\\\\
P17: If the actuator is up, and stays up, the extension sequence is never observed.\\\\
P18: The door manoeuvring from high to down or from down to high is made only if the gear is locked down or the gear is locked up (the door is manoeuvring means it is not open and it is not closed).\\\\
P19: Opening and closure door cannot happen in the same time.\\\\
P20: The full retraction and the full extension of the gear cannot happen in the same time.\\ \\
P21: Always, if the door manoeuvring or the gear manoeuvring, the pilot interface shows that the orange light is on.\\\\
P22: Always, if there is a failure in the door or in the gear, then the pilot interface shows that the red light is on.\\\\
P23: Always, if the gear is fully extended and locked down, and the door is locked in high position, then the pilot interface shows that the green light is on.\\\\
\noindent \textbf{Failures properties:}\\
P24: If the pilot push down the actuator and the door still locked in the closed position for more than 4s, then the door failure variable is set to true. \\\\
P25: If the actuator is pushed down and stays down, and the door still manoeuvring from high to down for more than 16s, then the door failure variable is set to true.\\\\
P26: If the door still manoeuvring from down to high for more than 44s after the actuator pushed down, then the door failure variable is set to true.\\   \\
P27: If the door seen not locked for more than 59s after the actuator pushed down, then the door failure variable is set to true.\\  \\
P28: In the landing mode, if the gear still locked in high position for more than 8s after the door opening, then the gear failure variable is set to true.\\    \\
P29: In the landing mode, if the gear seen not extended (still manoeuvring from high to down) for more than 20s after the door opening, then the gear failure variable is set to true.\\ \\
P30: In the landing mode, if the gear still not locked in down position for more than 24s from the door opening , then the gear failure variable is set to true.\\   \\
P31: After the take-off, in the retraction sequence, if the gear seen unlocked in down position for more than 8s after the door opening, then the gear failure variable is set to true.\\    \\
P32: After the take-off, in the retraction sequence, if the gear is not retracted (still manoeuvring from down to high) for more than 24s after the door opening, then the gear failure variable is set to true.\\         \\
P33: After the take-off, in the retraction sequence, if the gear is not locked in high position for more than 28s, then the gear failure variable is set to true.\\ 

\noindent\textbf{/* (2) Simple properties $(P_i*)$*/}\\
P1*: The door is completely open.\\
P2*: The door is not completely open.\\
P3*: The door is not closed.\\
P4*: The door is closed.\\
P5*: The door is locked.\\
P6*: The door is unlocked.\\
P7*: The door is manoeuvring from high to down.\\
P8*: The door is not manoeuvring from high to down.\\
P9*: The door is manoeuvring from down to high.\\ 
P10*: The door is not manoeuvring from down to high.\\ 
P11*: There is a failure in the door.\\
P12*: There is no failure in the door.\\
P13*: There is a failure in the gear.\\
P14*: There is no failure in the gear.\\
P15*: The gear is locked in high position.\\
P16*: The gear is not locked in high position.\\
P17*: The gear is locked in down position.\\
P18*: The gear is not locked in down position.\\
P19*: The gear is manoeuvring from high to down position.\\
P20*: The gear is not manoeuvring from high to down position.\\
P21*: The gear is manoeuvring from down to high position.\\
P22*: The gear is not manoeuvring from down to high position.\\
P23*: The gear is fully retracted.\\
P24*: The gear is not fully retracted.\\
P25*: The gear is fully extended.\\
P26*: The gear is not fully extended.\\
P27*: There is no light on in the pilot interface.\\
P28*: The red light is on in the pilot interface.\\   
P29*: The orange light is on in the pilot interface.\\ 
P30*: The green light is on in the pilot interface.\\
P31*: The actuator position is up.\\
P32*: The actuator position is down.\\ \\

\noindent
\textbf{Step 2 \textit{(modeling)}.} Formalisation of the required global properties with PSL.\\ \\
	
\begin{figure}[h]
		\begin{center}
			\makebox[\textwidth][c]{\includegraphics[scale=0.55]{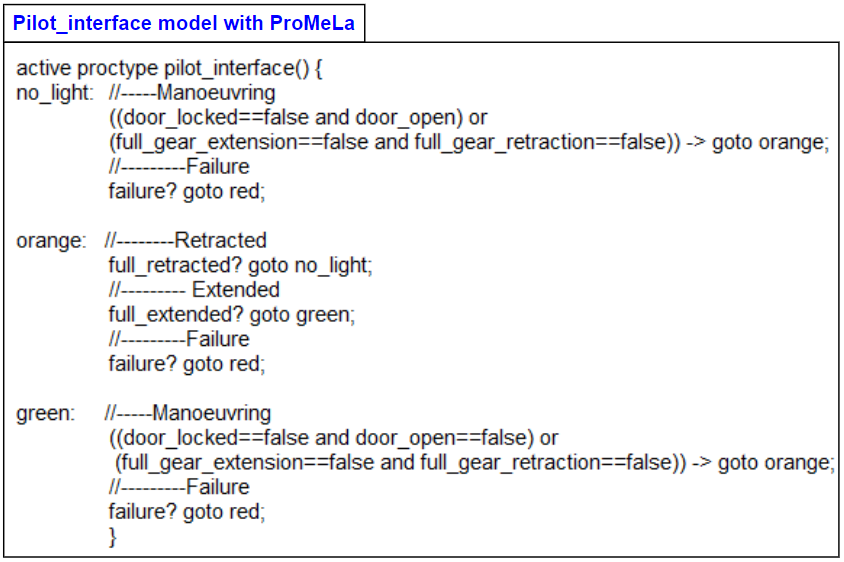}}  
			\caption{The pilot interface model with ProMeLa}
			\label{figure:interface_automata}
		\end{center}
	\end{figure}

\begin{figure}[h]
	\begin{center}
		\makebox[\textwidth][c]{\includegraphics[scale=0.50]{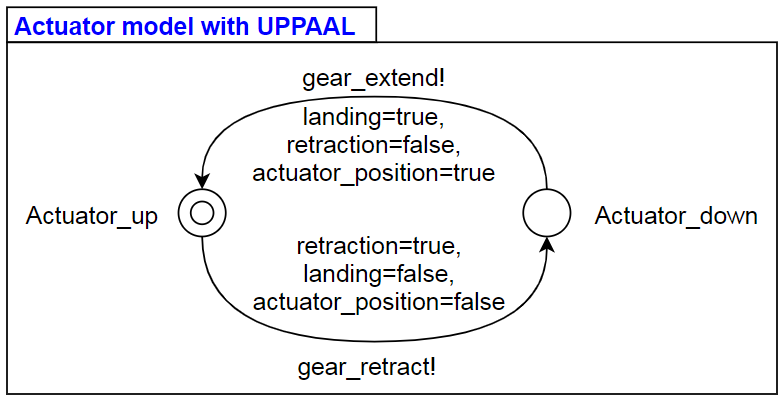}}  
		\caption{The actuator model with UPPAAL}
		\label{figure:actuator_automata}
	\end{center}	
\end{figure}

\begin{figure}[H]
	\begin{center}
		\makebox[\textwidth][c]{\includegraphics[scale=0.52]{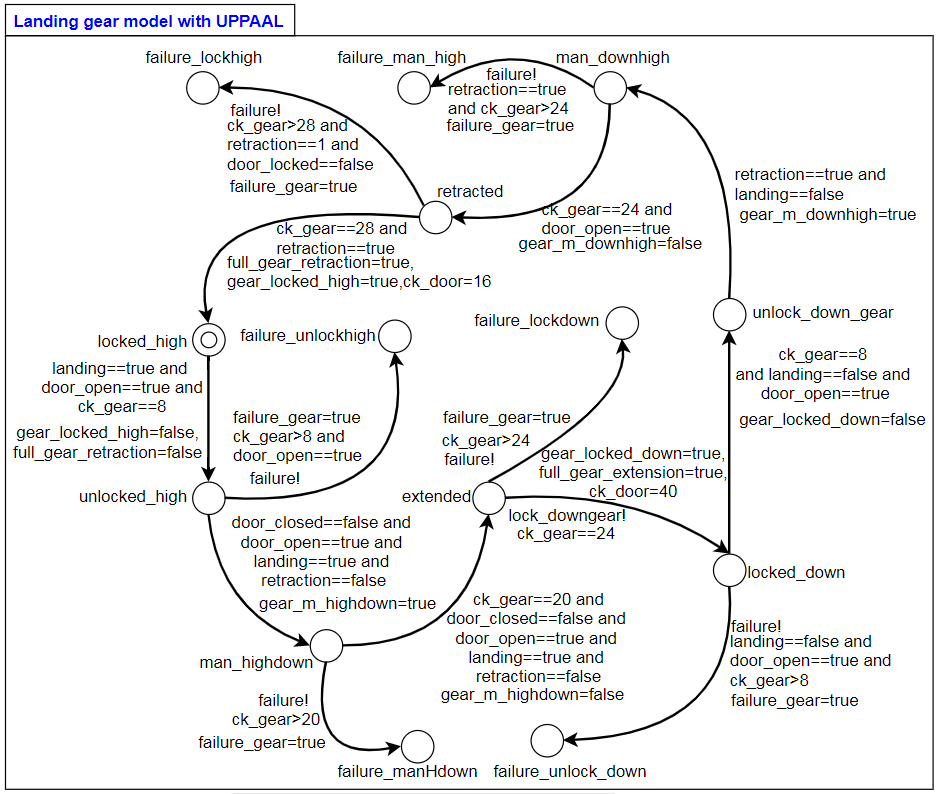}}  
		\caption{The landing gear model with UPPAAL}
		\label{figure:gear_automata}
	\end{center}
\end{figure}

\begin{figure}[H]
	\begin{center}
		\makebox[\textwidth][c]{\includegraphics[scale=0.40]{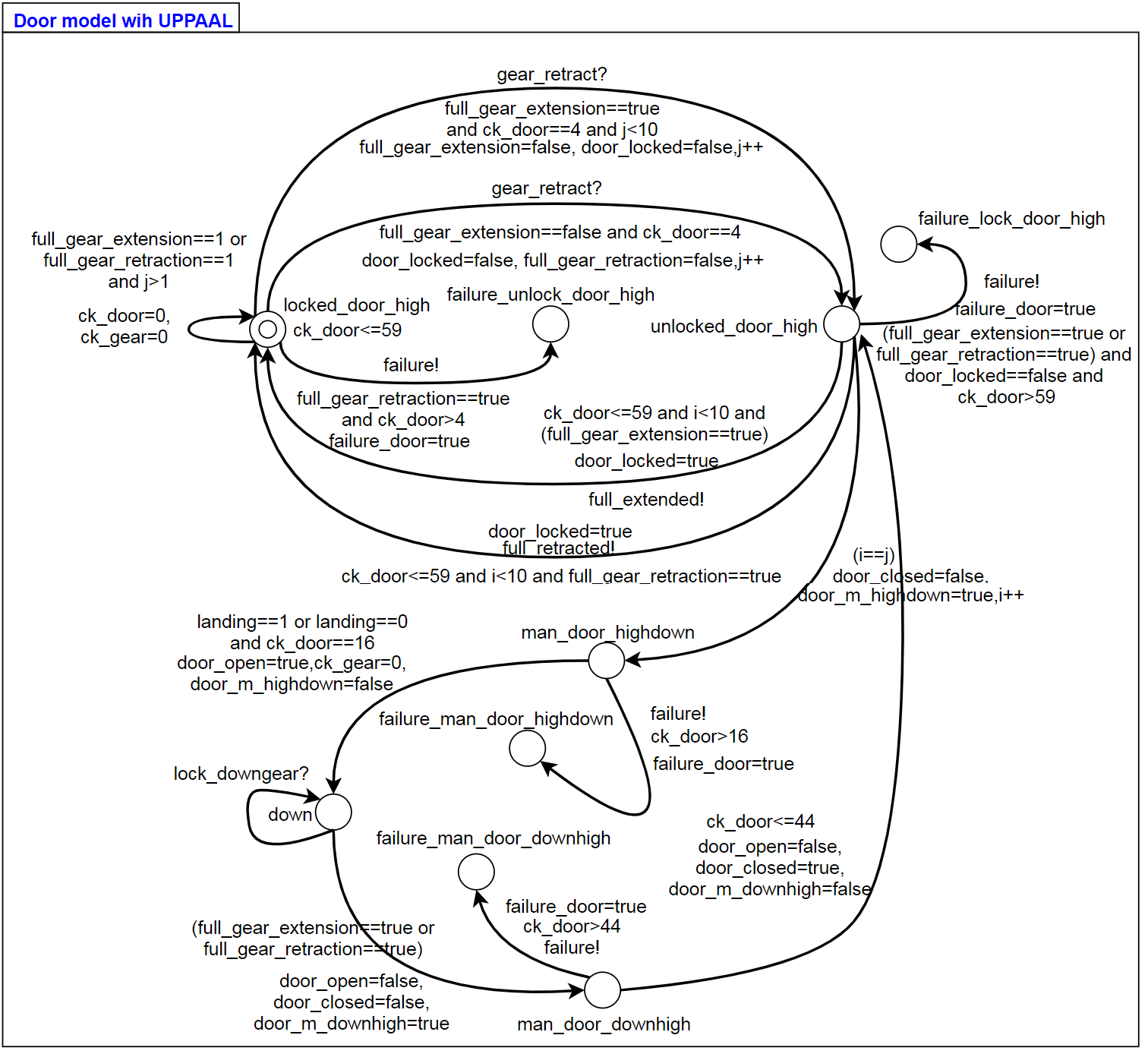}}  
		\caption{The door model with UPPAAL}
		\label{figure:door_automata}
	\end{center}
\end{figure}

\begin{tcolorbox}[enhanced,
	breakable,
	]
\texttt{ \begin{center}	\textcolor{olive}{\textbf{/*Formalized properties with PSL */}}\end{center} 
\noindent P1 : landing==true \barrow \evt full\_gear\_extension==true \e \\ gear\_locked\_down==true;\\\\  
P2 : \ag landing==false \e retraction==false \barrow \\full\_gear\_retraction==true \e door\_locked==true;\\\\
P2.1 : \ag landing==false \e retraction==false  \barrow  \\full\_gear\_retraction==true;\\\\
P2.2: \ag landing==false \e retraction==false\barrow door\_locked==true; \\\\
P3 : \ag gear.man\_highdown  \barrow   door\_open==true;\\\\
P4 : \ag gear.man\_downhigh  \barrow  door\_open==true;\\\\
P5 :\eg actuator\_position==true\barrow door\_closed==false \e  ck\_door==4;\\\\
P6 : \eg actuator\_position==true \barrow  door\_open==true \e ck\_door==16;\\\\
P7 : \ag gear.man\_highdown  \barrow  !door\_open==false;\\P8 : \ag gear.man\_downhigh  \barrow  !door\_open==false;\\\\
P9 : \eg actuator\_position==true \e door\_open==true   \barrow \\ gear\_locked\_high==false \e ck\_gear==8;\\\\
P10 : \eg actuator\_position==true \e  door\_open==true \e\\ gear\_locked\_high==false  \barrow  gear.extended \e ck\_gear==20;\\\\
P11 : \eg actuator\_position==true \e gear.extended   \barrow \\  gear\_locked\_down==true \e ck\_gear==24;\\\\
P12 : \eg actuator\_position==true \e gear\_locked\_down==true   \barrow \\  door\_closed==true \e door\_locked==true \e ck\_door==55;\\\\
P13 : \ag actuator\_position==false \e retraction==true \e\\ full\_gear\_extension==true \e\\ door\_closed==true \barrow door\_open==true \e ck\_door==16;\\\\
P14 : \ag actuator\_position==false \e retraction==true \e\\ door\_open==true \e full\_gear\_extension==true  \barrow \\ full\_gear\_retraction==true \e ck\_gear==20;\\\\
P15 : \eg retraction==true \e actuator\_position==false \e\\ full\_gear\_retraction==true  \barrow  door\_locked \e ck\_door==59;\\\\
P16 : \ag actuator\_position==true \e landing==true  \barrow \\ !retraction==true;\\\\
P17 : \ag actuator\_position==false \e retraction==true  \barrow \\!landing==true;\\\\
P18 : \ag door\_m\_highdown==true \orr door\_m\_downhigh==true   \barrow \\ gear\_locked\_down==true \orr gear\_locked\_high==true;\\\\
P19 : \ag door\_open==true  \barrow  door\_closed==false;\\\\
P20 : \ag full\_gear\_retraction==true  \barrow  full\_gear\_extension==false;\\ \\
P21 : \ag (door\_m\_downhigh==true \orr door\_m\_highdown==true) \e\\ (gear\_m\_highdown==true \orr\\ gear\_m\_downhigh==true)  \barrow  interface.orange;\\\\
P22 : \ag failure\_door==true \orr failure\_gear==true  \barrow interface.red;\\\\
P23 : \ag full\_gear\_extension==true \e gear\_locked\_down==true \e\\ door\_locked==true  \barrow  interface.green;
\begin{center}\textcolor{olive}{/*Failures Properties*/}\end{center}
P24 : \ef actuator.down \e door\_locked==true \e\\ door\_closed==true \e ck\_door>4  \barrow  failure\_door==true;\\\\
P25 : \ef actuator.down \e door\_m\_highdown==true \e ck\_door>16  \barrow \\  failure\_door==true;\\\\
P26 : \ef actuator.down \e door\_m\_downhigh==true \e ck\_door>44  \barrow \\  failure\_door==true;\\	\\   
P27 : \ef door\_locked==false \e actuator.down \e \\ck\_door>59  \barrow   failure\_door==true;\\\\
P28 : \ef landing==true \e gear\_locked\_high==true \e\\ door\_open==true \e ck\_gear>8  \barrow  failure\_gear==true;\\\\
P29 : \ef landing==true  \e gear\_m\_highdown==true \e\\ door\_open==true \e ck\_gear>20  \barrow  failure\_gear==true;\\\\
P30 : \ef landing==true \e gear\_locked\_down==false \e\\ door\_open==true \e ck\_gear>24  \barrow  failure\_gear==true;\\\\
P31:\ef retraction==true \e gear\_locked\_down==false \e\\ door\_open==true \e ck\_gear>8  \barrow  failure\_gear==true;\\\\
P32 : \ef retraction==true \e gear\_m\_downhigh==true \e\\ door\_open==true \e ck\_gear>24  \barrow  failure\_gear==true;\\\\
P33:\ef retraction==true \e gear\_locked\_high==false \e ck\_gear>28  \barrow  failure\_gear==true;\\\\
\textcolor{olive}{/*The variables that we need*/}\\
\textcolor{olive}{/*\textit{boolean variables}*/} \\
P341 : \bool door\_locked;\\
P342 : \bool door\_open;\\
P343 : \bool door\_closed; \\
P344 : \bool door\_m\_highdown; \\
P345 : \bool door\_m\_downhigh; \\
P346 : \bool failure\_door; \\
\\
P351 : \bool landing;\\
P352 : \bool retraction;\\
P353 : \bool actuator\_position;\\ 
\\
P361 : \bool full\_gear\_extension;\\
P362 : \bool full\_gear\_retraction; \\
P363 : \bool gear\_locked\_high; \\
P364 : \bool gear\_locked\_down;\\
P365 : \bool gear\_m\_highdown;\\
P367 : \bool gear\_m\_downhigh;\\
P368 : \bool failure\_gear;\\
\\
\textcolor{olive}{/*\textit{channel variables}*/}\\
\textcolor{olive}{/*The channel type used in UPPAAL model-checker is a very specific type, that is why we represent it like a boolean array.*/}\\ 
P371 : \bool gear\_extend[]; \\
P372 : \bool gear\_retract[];\\
P373 : \bool extend\_gear\_now[]; \\
P374 : \bool retract\_gear\_now[];\\
P375 : \bool lock\_highgear[];\\ 
P376 : \bool unlock\_gear[];\\
P378 : \bool failure[];\\
P379 : \bool full\_extended[]; \\
P3710 : \bool full\_retracted[];\\
P3711 : \bool lock\_downgear[];\\
\\
\textcolor{olive}{/*\textit{Clock variables}*/}\\
P38 : \clock ck\_door;\\
\\
P39 : \clock ck\_gear;\\
\\
\textcolor{olive}{/*\textit{Integer variables}*/}\\
P40 : \integer j;\\
\\
P41 : \integer i;	}
\end{tcolorbox}\leavevmode \\
\noindent\textbf{Step 3 \textit{(modeling)}.} We use UPPAAL and ProMeLa to model the components $interface$, $door$, $gear$, $actuator$. The models are depicted in the Figures \ref{figure:interface_automata}, \ref{figure:actuator_automata}, \ref{figure:gear_automata}, \ref{figure:door_automata}. The goal here is to show that we could select pre-modelled components even if they are in different languages and in different environments.  \\

	\noindent\textbf{Step 4 \textit{(modeling M$_{4}$)}.} We decompose the global properties with respect to the facets that we considered (data, safety, liveness, functionality, time); we obtain the \textit{generalized contract} decomposed with the facets. This will emphasize the concern to be dealt with later on.\\

\begin{tcolorbox}[enhanced,
	breakable,
	]
	\texttt{ \begin{center}	\textcolor{olive}{\textbf{/*Formalized and faceted properties with PSL */}}\end{center} 
	\textcolor{olive}{\textbf{/*Safety facet*/}}\\
	  P2 : \ag landing==false \e retraction==false \barrow \\full\_gear\_retraction==true \e door\_locked==true;\\\\
	  P2.1 : \ag landing==false \e retraction==false \barrow \\full\_gear\_retraction==true;\\\\
	  P2.2 : \ag landing==false \e retraction==false \barrow \\ door\_locked==true; \\\\
	  P3 : \ag gear.man\_highdown \barrow door\_open==true;\\\\
	  P4 : \ag gear.man\_downhigh  \barrow  door\_open==true;\\\\
	  P7 : \ag gear.man\_highdown  \barrow  !door\_open==false;\\\\
	  P8 : \ag gear.man\_downhigh  \barrow  !door\_open==false;\\\\
	  P13 : \ag actuator\_position==false \e \\ retraction==true \e full\_gear\_extension==true \e\\ door\_closed==true \barrow door\_open==true \e ck\_door==16;\\\\
	  P14 : \ag actuator\_position==false \e\\ retraction==true \e door\_open==true \e full\_gear\_extension==true  \barrow  full\_gear\_retraction==true \e ck\_gear==20;\\\\
	  P16 : \ag actuator\_position==true \e landing==true  \barrow  !retraction==true;\\\\
	  P17 : \ag actuator\_position==false\\ \e retraction==true  \barrow  !landing==true;\\\\
	  P18 : \ag door\_m\_highdown==true \orr \\door\_m\_downhigh==true  \barrow  gear\_locked\_down==true \orr \\gear\_locked\_high==true;\\\\
	  P19 : \ag door\_open==true  \barrow  door\_closed==false;\\\\
	  P20 : \ag full\_gear\_retraction==true  \barrow  \\full\_gear\_extension==false; }
\end{tcolorbox}

\begin{tcolorbox}[enhanced,
	breakable,
	]\texttt{
	\textcolor{olive}{\textbf{/*Liveness facet*/}}\\
	\noindent  P1 : \always (landing==true  \barrow \evt\\ full\_gear\_extension==true \e gear\_locked\_down==true);\\ \\
	  P5 : \always (actuator\_position==true  \barrow \evt  \\door\_closed==false \e  ck\_door==4);\\\\
	  P6 : \always (actuator\_position==true   \barrow \evt \\ door\_open==true \e ck\_door==16);\\\\
	  P9 : \always (actuator\_position==true \e  door\_open==true \\   \barrow \evt gear\_locked\_high==false \e ck\_gear==8);\\\\
	  P10 : \always (actuator\_position==true \e  door\_open==true \e gear\_locked\_high==false  \barrow \evt gear.extended \e \\ck\_gear==20);\\\\
	  P11 : \always (actuator\_position==true \e gear.extended\\ \barrow \evt  gear\_locked\_down==true \e ck\_gear==24);\\\\
	  P12 : \always (actuator\_position==true \e\\ gear\_locked\_down==true \barrow \evt  door\_closed==true \e door\_locked==true \e ck\_door==55);\\}
\end{tcolorbox}

\begin{tcolorbox}[enhanced,
	breakable,
	]\texttt{
   	\textcolor{olive}{\textbf{/*Functionality facet*/}}\\
   	  P15 : \eg retraction==true \e  \\actuator\_position==false \e full\_gear\_retraction==true  \barrow  \\door\_locked \e ck\_door==59;\\\\
	  P21 : \ag (door\_m\_downhigh==true \orr\\ door\_m\_highdown==true) \e (gear\_m\_highdown==true \orr\\ gear\_m\_downhigh==true)  \barrow  interface.orange;\\\\
	  P22 : \ag failure\_door==true \orr failure\_gear==true  \barrow  interface.red;\\\\
	  P23 : \ag full\_gear\_extension==true \e\\ gear\_locked\_down==true \e door\_locked==true  \barrow  interface.green;\\}
\end{tcolorbox}

\begin{tcolorbox}[enhanced,
	breakable,
	]\texttt{
	\textcolor{olive}{\textbf{/*Time facet*/}}\\
	  P24 : \ef actuator.down \e door\_locked==true \e\\ door\_closed==true \e ck\_door>4  \barrow  failure\_door==true;\\\\
	  P25 : \ef actuator.down \e door\_m\_highdown==true \e ck\_door>16  \barrow   failure\_door==true;\\\\
	  P26 : \ef actuator.down \e door\_m\_downhigh==true \e ck\_door>44  \barrow   failure\_door==true;\\	\\						    
	  P27 : \ef door\_locked==false \e\\ actuator.down \e ck\_door>59  \barrow   failure\_door==true;\\\\
	  P28 : \ef landing==true \e gear\_locked\_high==true \e door\_open==true \e ck\_gear>8  \barrow  failure\_gear==true;\\\\
	  P29 : \ef landing==true  \e gear\_m\_highdown==true \e door\_open==true \e ck\_gear>20  \barrow  failure\_gear==true;\\\\
	  P30 : \ef landing==true \e gear\_locked\_down==false \e door\_open==true \e ck\_gear>24  \barrow  failure\_gear==true;\\\\
	  P31 : \ef retraction==true \e gear\_locked\_down==false \e door\_open==true \e ck\_gear>8  \barrow  failure\_gear==true;\\\\
	  P32 : \ef retraction==true \e gear\_m\_downhigh==true \e door\_open==true \e ck\_gear>24  \barrow  failure\_gear==true;\\\\
	  P33 :\ef retraction==true \e gear\_locked\_high==false \e ck\_gear>28  \barrow  failure\_gear==true;\\}
\end{tcolorbox}

\begin{tcolorbox}[enhanced,
	breakable,
	]\texttt{
\textcolor{olive}{\textbf{/*Data facet*/}}\\
\textcolor{olive}{/*The variables that we need*/}\\
\textcolor{olive}{/*\textit{boolean variables}*/} \\
P341 : \bool door\_locked;\\
P342 : \bool door\_open;\\
P343 : \bool door\_closed; \\
P344 : \bool door\_m\_highdown; \\
P345 : \bool door\_m\_downhigh; \\
P346 : \bool failure\_door; \\
\\
P351 : \bool landing;\\
P352 : \bool retraction;\\
P353 : \bool actuator\_position;\\ 
\\
P361 : \bool full\_gear\_extension;\\
P362 : \bool full\_gear\_retraction; \\
P363 : \bool gear\_locked\_high; \\
P364 : \bool gear\_locked\_down;\\
P365 : \bool gear\_m\_highdown;\\
P367 : \bool gear\_m\_downhigh;\\
P368 : \bool failure\_gear;\\
\\
\textcolor{olive}{/*\textit{channel variables}*/}\\
\textcolor{olive}{/*The channel type used in UPPAAL model-checker is a very specific type, that is why we represent it like a boolean array.*/}\\ 
P371 : \bool gear\_extend[]; \\
P372 : \bool gear\_retract[];\\
P373 : \bool extend\_gear\_now[]; \\
P374 : \bool retract\_gear\_now[];\\
P375 : \bool lock\_highgear[];\\ 
P376 : \bool unlock\_gear[];\\
P378 : \bool failure[];\\
P379 : \bool full\_extended[]; \\
P3710 : \bool full\_retracted[];\\
P3711 : \bool lock\_downgear[];\\
\\
\textcolor{olive}{/*\textit{Clock variables}*/}\\
P38 : \clock ck\_door;\\
\\
P39 : \clock ck\_gear;\\
\\
\textcolor{olive}{/*\textit{Integer variables}*/}\\
P40 : \integer j;\\
\\
P41 : \integer i;\\}
\end{tcolorbox}	

\begin{figure}[H]
	\centering
	\makebox[\textwidth][c]{\includegraphics[scale=0.6]{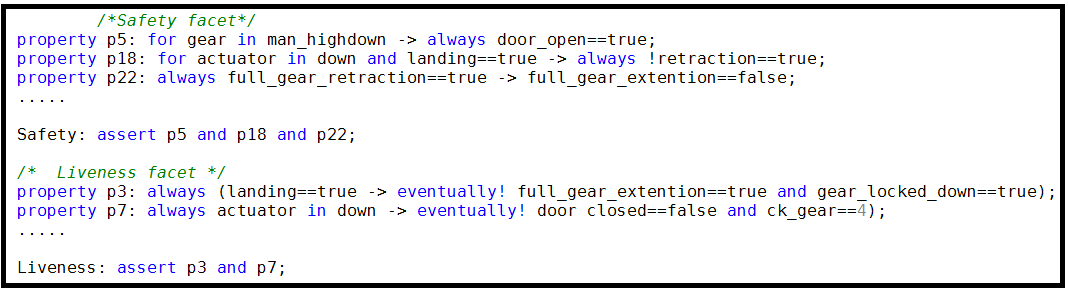}}
	\captionsetup{justification=centering}\caption{A part of structured and faceted properties }
	\label{figure:structured_properties}
\end{figure}

	\noindent
\textbf{Step 5 \textit{(modeling M$_{5}$)}.} We express the structured and formalized properties with the PSL language, as depicted in figure \ref{figure:structured_properties}. \\
\begin{tcolorbox}[enhanced,
	breakable,
	]
	\texttt{\textcolor{olive}{\textbf{/*\textit{Safety facet}*/}}\\
\propertyb P3: \for gear \inn man\_highdown \barrow  \xspace \always door\_open==true;\\
\propertyb P16: \for actuator \inn down \e landing==true \barrow \always !retraction==false;\\
\propertyb P20: \always full\_gear\_retraction==true \barrow full\_gear\_extension==false;\\
$\cdots$\\ 
Safety: \assert P3 \e P16 \e P20; }
		
\end{tcolorbox}
\begin{tcolorbox}[enhanced,
	breakable,
	]
	\texttt{\textcolor{olive}{\textbf{/*\textit{Liveness facet}*/}}\\
		\propertyb P1: \always (landing==true \barrow \xspace \evt full\_gear\_extension==true \e gear\_locked\_down==true);\\
		\propertyb P5: \always actuator \inn down \barrow \evt door\_closed==false \e ck\_gear==4);\\
		$\cdots$\\ 
		Safety: \assert P1 \e P5; }
	
\end{tcolorbox}

\begin{figure}[H]
	\centering
	\makebox[\textwidth][c]{\includegraphics[scale=0.42]{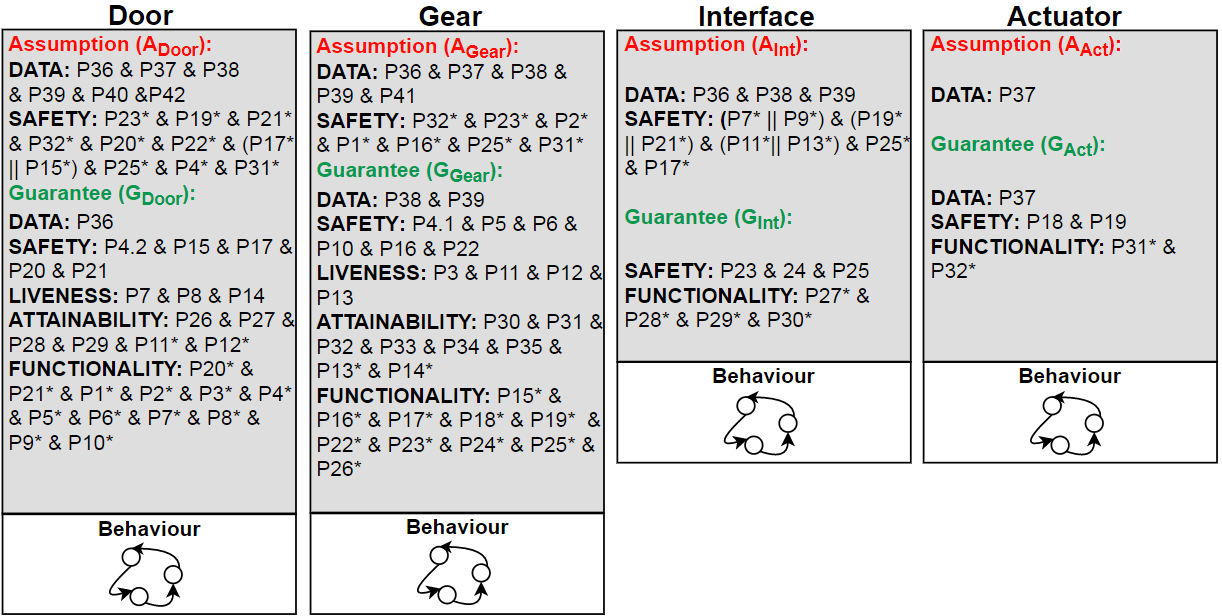}}
	\caption{Normalized components }
	\label{figure:normalized_components}
\end{figure}

\noindent
\textbf{Step 6 \textit{(modeling M$_{6}$)}.} Normalizing the individual components \cite{minarets}. We integrate the assumptions and guarantees of each individual component. Figure \ref{figure:normalized_components} shows the normalized individual components.
\noindent
\textbf{Step 7 \textit{(modeling M$_{7}$)}.} According to the result of \textit{Step 4}, we may add a facet of the global property to a component, or ignore some of its facets, if necessary. In the current case study it is not necessary to add or ignore a facet (see Figure \ref{figure:normalized_components}).\\

\begin{figure}[H]
	\centering
	 \makebox[\textwidth][c]{\includegraphics[scale=0.42]{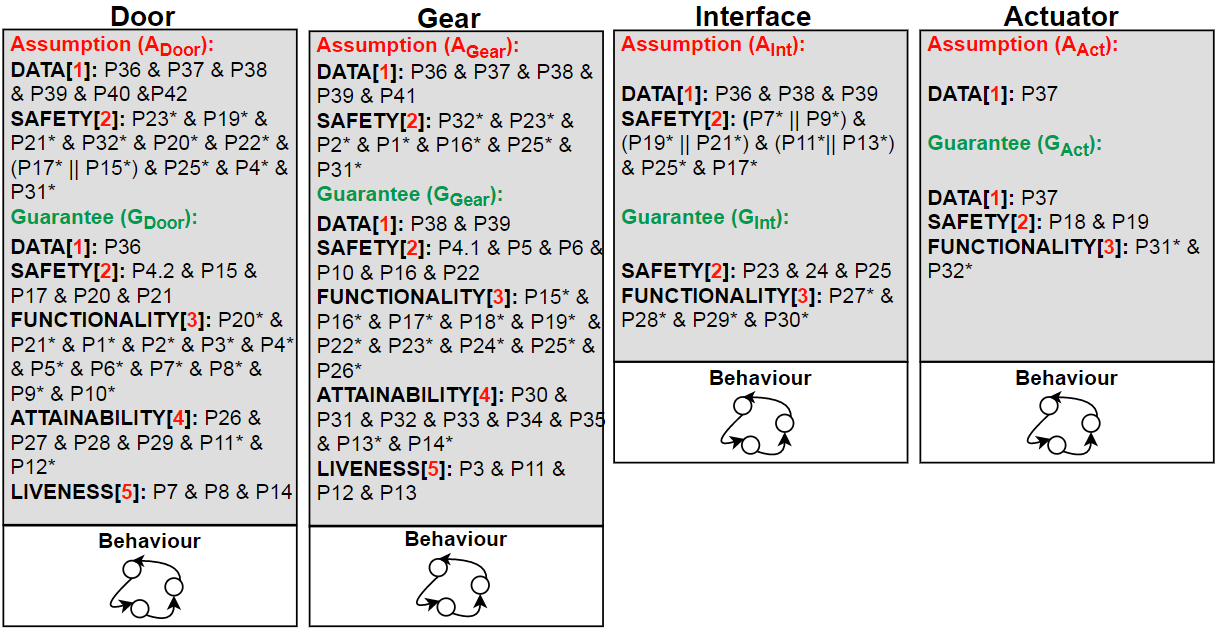}}
	\captionsetup{justification=centering}\caption{Normalized components with the prioritized facets}
	\label{figure:normalized_components_priority}
\end{figure}

\noindent
\textbf{Step 8 \textit{(modeling M$_{8}$)}.} We attribute the following priorities to each facet (data=1, safety=2, functionality=3 time=4, liveness=5); were 1 is the highest priority). We obtain ordered layers with respect to facets and properties as well. The order of layer is mentioned in Figure \ref{figure:normalized_components_priority}. The verification by layer allows one to verify the contracts by priority; from a very important layer (primary) to a less important layer (secondary). If the behaviour of our system does not satisfy a primary layer of contract, then, it is not necessary to continue the verification with the other layers.\\

\noindent
\textbf{Step 9 \textit{(verification V$_{1}$)}.} We have to check the appropriate functioning of each normalized individual component if tools exist for that, if the language supports all the needed facets and if the required data are available. As we use ProMeLa we have an adequate tool (SPIN) but, the only component $interface$ modelled with ProMeLa cannot be verified without composition with its environment; then, a translation to the targeted tool is mandatory.\\

\begin{figure}[H]
	\centering
	\makebox[\textwidth][c]{\includegraphics[scale=0.60]{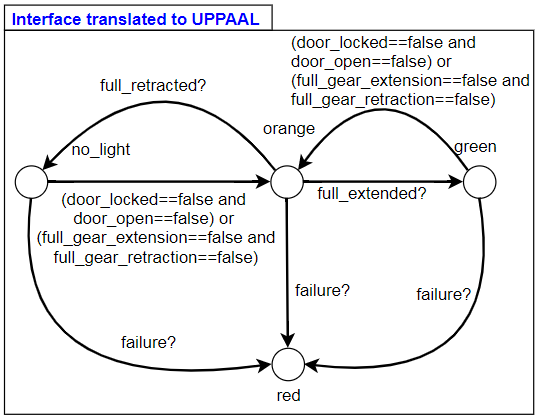}}
	\captionsetup{justification=centering}\caption{Pilot interface component translated from ProMeLa to UPPAAL}
	\label{figure:interface_uppaal}
\end{figure}

\noindent
\textbf{Step 10 \textit{(modeling M$_{9}$)}.} Translation of the ProMeLa component $Interface$ to the targeted tool UPPAAL using the algorithm presented in our RR \cite{minarets}, we obtain a component $Interface$ ready for composition (see Figure \ref{figure:interface_uppaal}).\\
\noindent
\textbf{Step 11 \textit{(modeling M$_{10}$)}.} 
 We compose the generalized contracts of the components (facet by facet). The Figure below shows the composition of the generalized contracts of the components $interface$ and $actuator$ to build the $cockpit$ component.\\
 \begin{figure}[H]
 	\centering
 	\makebox[\textwidth][c]{\includegraphics[scale=0.450]{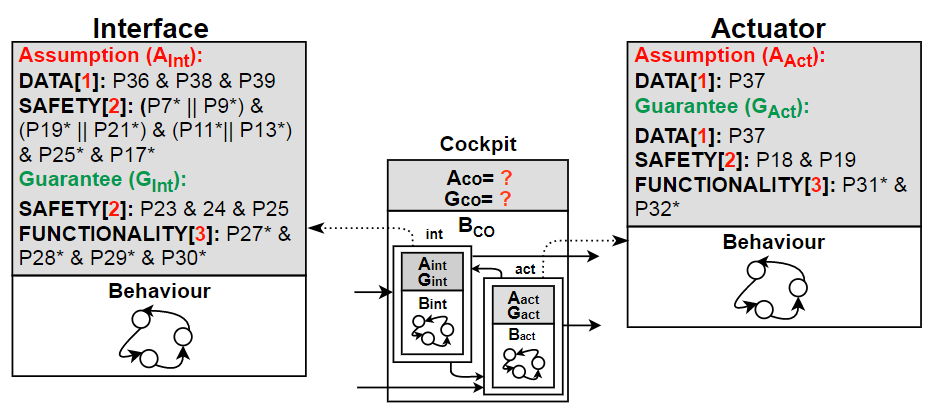}}
 	\captionsetup{justification=centering}\caption{The composition of the generalized contracts of the components Interface and Actuator }
 	\label{figure:composition_GC}
 \end{figure}
As depicted in the Figure \ref{figure:composition_GC}, the generalized contracts $GC$ of the component $Interface$, $GC_{Int}=(A_{Int}, G_{Int})$ is:\\ \\
$A_{Int}=\\
\{(\textbf{DATA}[1]:\ P34\ \&\ P36\ \&\ P37),\\
 \hspace*{0.2cm}(\textbf{SAFETY}[2]:\ (P5^*\ ||\ P7^*)\ \&\ (P17^*\ ||\ P19^*)\ \&\ (P9^*\ ||\ P11^*)\ \&\ P23^*\ \&\ P15^*)\}$\\
 
\noindent$G_{Int}=\\
\{(\textbf{SAFETY}[2]:\ P21\ \&\ 24\ \&\ P23),\\
  \hspace*{0.2cm}(\textbf{FUNCTIONALITY}[3]:\ P25*\ \&\ P26*\ \&\ P27*\ \&\ P28*)\}$\\
   
\noindent The generalized contracts $GC$ of the component $Actuator$, $GC_{Act}=(A_{Act}, G_{Act})$ is:\\\\
$A_{Act}=\\
\{(\textbf{DATA}[1]:\ P35)\}$\\

\noindent$G_{Act}=\\
\{(\textbf{DATA}[1]:\ P35),\\
\hspace*{0.2cm}(\textbf{SAFETY}[2]:\ P16\ \&\ P17),\\
\hspace*{0.2cm}(\textbf{FUNCTIONALITY}[3]:\ P29^*\ \&\ P30^*)\}$\\

\noindent Note that the components $Interface$ and $Actuator$ are in parallel and share variables:\\
$\textsf{\textit{vars}}(\textsf{\textit{ppty}}(G_{Int},FUNCTIONALITY)) \cap \textsf{\textit{vars}}(\textsf{\textit{ppty}}(G_{Act},FUNCTIONALITY))=\\
\{landing,\ ck\_gear,\ door\_open,\ failure\_gear\}$.\\ There is no inconsistency between the facets because in each facet we have:\\
The same value of the variables: $landing=true,\ door\_open=true,\ failure\_gear=true$ and in case of the variable $ck\_gear$ we have $ck\_gear>8$ $\Rightarrow$ $ck\_gear>20$ and $ck\_gear>8$ $\Rightarrow$ ck\_gear>24.\\\\ 
\noindent The composition of the generalized contract $GC_{Int} \oplus GC_{Act}$ is:\\ 

\noindent $A_{Cockpit}=\\
\{(\textbf{DATA}[1]:\ (P34\ \&\ P36\ \&\ P37)\boldmath\bigwedge (P35)),\\
   \hspace*{0.2cm}(\textbf{SAFETY}[2]:\ (P5^*\ ||\ P7^*)\ \&\ (P17^*\ ||\ P19^*)\ \&\ (P9^*\ ||\ P11^*)\ \&\ P23^*\ \&\ P15^*)\}$\\

\noindent$G_{Cockpit}=\\
\{(\textbf{SAFETY}[2]:\ (P21\ \&\ 24\ \&\ P23)\boldmath\bigwedge (P16\ \&\ P17)),\\
\hspace*{0.2cm}(\textbf{FUNCTIONALITY}[3]:\ (P25^*\ \&\ P26^*\ \&\ P27^*\ \&\ P28^*) \boldmath\bigwedge (P29^*\ \&\ P30^*))\}$\\\\



\begin{figure}[H]
	\begin{center}
		\includegraphics[scale=0.5]{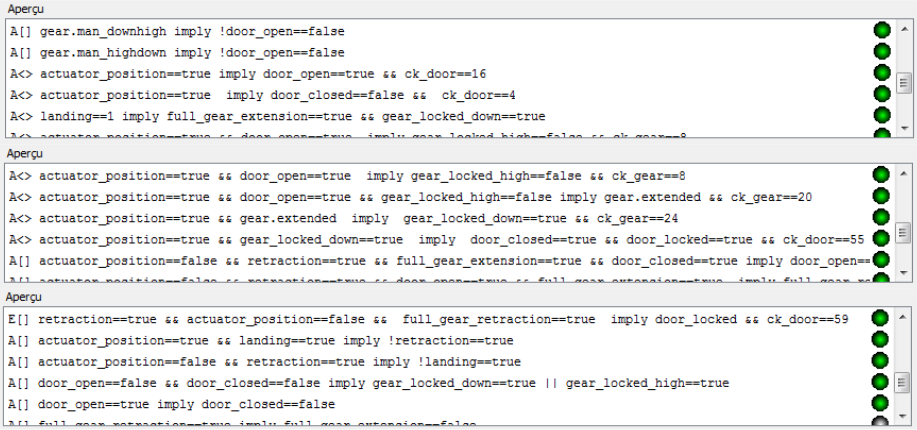}  
		\caption{Status of properties verification with UPPAAL }
		\label{figure:verification_UPPAAL}
	\end{center}
\end{figure}

\noindent \textbf{Step 12 \textit{(verification V$_{2}$)}.} We translate the \textit{generalized contract} of each individual component, from \textit{PSL} to the UPPAAL language. The following property is an extract of the translation from the RP component.\\\\
{
	\texttt{
		\hspace*{0.7cm}{A[] gear.man\_highdown imply !door\_open==false}  \\
		\hspace*{0.7cm}{A[] failure\_door==true or failure\_gear==true imply interface.red}\footnote{were "A [ ] Prop" denotes the "always property".}}}  \\\\
\textbf{Step 13 \textit{(verification V$_{3}$)}.} After the composition of the system in $step 11$, we verify the global property of the same layer together,facet by facet (data, security, time, functionality); also, the primary properties before the secondary properties. At the end of this step, we obtain the verified LGS system. Figure \ref{figure:verification_UPPAAL} shows the verification status of some translated properties.   

\section{Related work}\label{section:related work}
The landing gear system were widely presented by the software engineering community; some of these works mainly based on Event-B such as \cite{pascal_landing, laleau_landing, lukas_landing}. If we focus on the analysis of the various properties, we remark that the safety, time and liveness properties were formalized and proved, but from our point of view the analysis of time and liveness properties is heavy using Event-B; it is not possible to directly verify the properties of these facets, but it is necessary to improve Event-B method and enriching its tools to could analyse these facets. 

The LGS were also presented in the work \cite{landing_richard} using multi-machine hybrid Event-B, an hybrid modeling that combines software, hydraulic and mechanical parts; but without the verification steps. 

The work \cite{landing_paolo} presented LGS with different method, they used the Abstract State Machine (ASM) and ASMETA framework for modeling and analysis. They take into consideration only the weaker versions of properties without taking into consideration the real time aspects.

Frédéric Boniol in the work \cite{landing_boniol_review} made a benchmark for techniques and tools of some approaches that dealt with LGS, describe more the case study in \cite{landing_boniol_details} and made different experimentations in \cite{landing_boniol_experimentation} using different tools like UPPAAL, Lustre, SMV ...etc (the case study is made completely in each tool). The specification were translated manually from Esterel to Lustre and UPPAAL. In \cite{landing_boniol_experimentation} there were no compositional verification. In our work we compose the system using generalized contracts; after that, we verify the overall consistency facet by facet.

 In addition, unlike the most approaches that present the LGS, in our work we deal with heterogeneous components; our system is a combination of components modelled with different languages and in different environments, the component $interface$ is modelled with the SPIN tool with the ProMeLa language, and the components $Gear$, $Door$, $Actuator$ are modelled with UPPAAL, in the same case study we use some components with UPPAAL and other components with SPIN. Moreover, we specify the various properties with the PSL language.

\section{Conclusion}\label{section:conclusion}
 Our \textsf{Minarets} method for complex and heterogeneous systems modeling and analysis, emphasizes the stepwise composition of heterogeneous components through their generalized contracts.

 In this paper, We have shown how one can reduce the complexity of the global modeling and the global analysis of complex and heterogeneous systems; we experimented our approach with the well known case study of the landing gear system; it involves different facets (data, safety, liveness, functionality and time). The generalized contracts are first expressed in PSL. We normalized and composed the heterogeneous components. We have translated the properties into input languages of UPPAAL/SPIN model checkers, then checked them with respect to the various facets.
  We defined in a formal way the semantics of contracts composition, and checked without ambiguity the correctness of the overall LGS, facet by facet, using the generalized contracts principles.

In our future works, we will study deeply the interferences between facets and the compatibility between data types and different units during the composition; we will support our method with a comparison between the LTS of each used language in order to be able to choose a suitable and a global LTS that covers all facets; finally, we will propose a tool to guide the users in normalising, composing and verifying the heterogeneous components. This will furthermore strengthen the foundations of the proposed method, and enable the contract management tool construction.
\bibliographystyle{splncs04}

\bibliography{./biblioFile}
\end{document}